\documentclass[aps,epsfig,11pt,manuscript]{revtex4}
\usepackage{amsmath}
\usepackage{amssymb,graphicx}
\usepackage{epsfig}
\usepackage{graphicx}
\usepackage{enumerate}
\usepackage{graphicx,wrapfig}

\usepackage{color}

\makeatletter
\@ifundefined{textcolor}{}
{%
 \definecolor{BLACK}{gray}{0}
 \definecolor{WHITE}{gray}{1}
 \definecolor{RED}{rgb}{1,0,0}
 \definecolor{GREEN}{rgb}{0,1,0}
 \definecolor{BLUE}{rgb}{0,0,1}
 \definecolor{CYAN}{cmyk}{1,0,0,0}
 \definecolor{MAGENTA}{cmyk}{0,1,0,0}
 \definecolor{YELLOW}{cmyk}{0,0,1,0}
 }

\newcommand{\bea}{\begin{eqnarray}}
\newcommand{\eea}{\end{eqnarray}}

\newcommand{\be}{\begin{equation}}
\newcommand{\ee}{\end{equation}}

\begin{document}

\title
%
%
{Full density matrix dynamics for large quantum systems:
Interactions, Decoherence and Inelastic effects}
%
%
%
%

%
\author{Manas Kulkarni}
\address{Chemical Physics Theory Group, Department of Chemistry,
University of Toronto, 80 Saint George St. Toronto, Ontario, Canada
M5S 3H6} \address{Department of Physics, University of Toronto, 60
Saint George St. Toronto, Ontario, Canada M5S 1A7}
\author{Kunal L. Tiwari}
\address{Chemical Physics Theory Group, Department of Chemistry,
University of Toronto, 80 Saint George St. Toronto, Ontario, Canada
M5S 3H6}
\address{Department of Physics, University of Toronto, 60 Saint
George St. Toronto, Ontario, Canada M5S 1A7}
\author{Dvira Segal}
\address{Chemical Physics Theory Group, Department of Chemistry,
University of Toronto, 80 Saint George St. Toronto, Ontario, Canada
M5S 3H6} 


\date{\today}

\begin{abstract}
We develop analytical tools and numerical methods for time evolving
the {\it total} density matrix of the finite-size Anderson model.
The model is composed of two finite metal grains, each prepared in
canonical states of differing chemical potential and connected
through a single electronic level (quantum dot or impurity). Coulomb
interactions are either excluded all together, or allowed on the dot
only. We extend this basic model to emulate decoherring and
inelastic scattering processes for the dot electrons with the probe
technique. Three methods, originally developed to treat impurity
dynamics, are augmented to yield global system dynamics: the quantum
Langevin equation method, the well known fermionic trace formula,
and an iterative path integral approach. The latter accommodates
interactions on the dot in a numerically exact fashion. We apply the
developed techniques to two open topics in nonequilibrium many-body
physics: (i) We explore the role of many-body electron-electron
repulsion effects on the dynamics of the system. Results, obtained
using exact path integral simulations, are compared to mean-field
quantum Langevin equation predictions.
(ii) We analyze aspects of quantum equilibration and thermalization
in large quantum systems using the probe technique, mimicking
elastic-dephasing effects and inelastic interactions on the dot.
Here, unitary simulations based on the fermionic trace formula are
accompanied by quantum Langevin equation calculations.

\end{abstract}

\maketitle

\section{Introduction}

There has been recently a great deal of interest in simulating the
real-time dynamics of quantum systems, open or closed, prepared in a
nonequilibrium state \cite{Rev}. These investigations have been
spurred by recent experimental breakthroughs in the ability to watch
out-of-equilibrium dynamics, for example, in cold atomic gases
\cite{bloch}, or in on-chip superconducting circuits  \cite{houck}.
This endeavor is fundamentally important for resolving basic issues
in quantum dynamics, and in particular, for understanding
equilibration and thermalization in quantum systems \cite{Rev}.
%
The nonequilibrium dynamics of the eminent Anderson model
\cite{anderson}, composed of a single electronic level (quantum dot)
coupled to two metals, has in particular been of great interest.
This is because it is perhaps the simplest platform for probing both
equilibrium and out-of-equilibrium physics in a many-body system.
The model is integrable, even when electron-electron (e-e) repulsion
effects are accounted for on the dot, and its integrability has been
exploited for resolving its transport behavior \cite{rmk}.

A central strategy in most analytic and numerical tools devoted to the
Anderson model, and impurity systems at large, is the separation of the
total system into a subsystem (dot) and the environment (metals,
referred to as reservoirs). The latter are typically assumed to be
infinite-dissipative and are maintained in one of the canonical ensembles
of statistical mechanics. This assumption allows one to treat the
effect of the reservoirs on the subsystem within a self-energy term.
However, once the reservoirs are traced out, one cannot describe their
explicit dynamics. Among the numerical approaches
developed along these lines we list the time-dependent numerical
renormalization-group method \cite{anders}, real-time diagrammatic
Monte Carlo techniques \cite{marco} and path integral approaches
\cite{egger1,egger2}. These methods place
focus on quantities such as the dot occupancy, transmission
probability, conductance, current, noise, and correlations on the
{\it impurity}. The dynamics of the {\it total} system, including the
electron reservoirs, has not yet been explored since general tools
for simulating the overall dynamics in a system-bath scenario are
still missing.

\begin{figure}[htbp]
 {\hbox{\hspace{15mm}\epsfxsize=120mm \epsffile{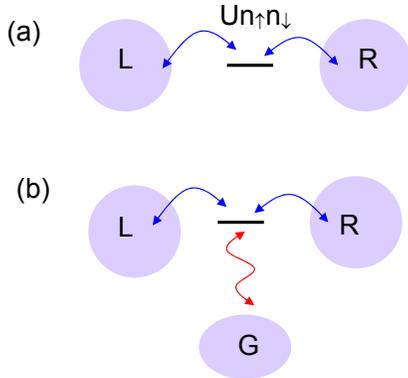}}} \vspace{-105mm}\caption{
Two metallic grains (reservoirs) separately prepared in a grand
canonical- diagonal state. At $t_0$ the reservoirs are put into
contact through a single electronic state.
We study three variants of this systems: a case without interactions,
(a)  allowing for electron repulsion effects on the dot only, and (b)
accommodating decoherring
and inelastic effects on the dot, by coupling its electrons to a $G$
reservoir, serving as a dephasing or a voltage probe.} \label{FigS}
\end{figure}

The current work develops analytical and numerical treatments of
global system evolution based on established impurity dynamics
techniques. These tools allow investigation of the roles of e-e
interactions and decoherence and dissipation effects on
nonequilibrium reservoirs dynamics. We focus on the finite-size
Anderson model composed of two metallic grains weakly coupled
through a single electronic level. We refer to the metal grains,
each composed of  $N\sim 100-500$ electronic states and $n\sim
50-200$ electrons as ``reservoirs" alluding to their high density of
states (DOS). This large DOS allows the reservoirs' effect on the
dot (subsystem) to be absorbed into a positive real self-energy
function lending to a quantum Langevin equation (QLE) description
\cite{ford,sen,DharDM}, as we explain below. A schematic representation
is presented in Fig. \ref{FigS}. We are interested in following the
real-time dynamics of {\it both} the dot and the reservoirs degrees
of freedom. As an initial condition, we assume that each reservoir
is prepared in a distinct Gibbs-like grand canonical state at a
different chemical potential but at the same temperature.

In the absence of dephasing and inelastic effects, the dynamics of
the total density matrix is followed by extending three approaches:
(i) the quantum Langevin equation method \cite{ford,sen,DharDM},
adopted here both in the noninteracting limit and in the mean-field
(MF) regime, (ii) fermionic trace formula \cite{israel}, used here
for simulating the exact dynamics of the noninteracting model, and
(iii) an influence functional path integral method
\cite{dvira_pccp,dvira_prb}, employed to treat interactions beyond
the perturbative regime. In the latter half of the paper, these
techniques are used to study reservoir population evolution both
without and with Coulomb repulsion effects on the dot, and in the
presence of emulated dephasing and inelastic scattering effects.

While Coulomb interactions are explicitly introduced here, the
inclusion of dephasing and inelastic effects warrants further
discussion. The origin of such processes are many-body interactions
in the system, e.g., electron phonon coupling. Since an explicit and
exact inclusion of these interactions is extremely challenging
\cite{Eran,Thoss,Thorwart,Andrei}, phenomenological techniques have
been developed in their stand, \cite{Buttiker1,Buttiker2,Beenakker}.
In the case of elastic decoherring processes the technique is
referred to as a ``dephasing probe". In the case of inelastic
scattering processes, it is referred to as a ``voltage probe". These
probes are electron reservoirs, prepared such that, there is no
either energy resolved or total net electron flow from the
$L$-dot-$R$ system towards these probes. For a scheme of this model,
see Fig. \ref{FigS}(b). It should be noted that elastic-decoherring
processes or inelastic effects are only {\it emulated} here by the
probes. The overall dynamics can be still simulated using the
unitary trace formula technique \cite{qle}. We also extend the QLE
method to include a probe, and time evolve the system.
%
Since our calculations provide the real-time dynamics of the {\it
full} density matrix (DM), the process of equilibration and
thermalization in a finite quantum system can
now be studied \cite{h1,h2,barra,qle}. 
Particularly, we find that when only decoherence
effects are allowed, the system approaches a non-canonical
equilibrium state. In contrast, when inelastic processes are
included, the reservoirs relax towards a common Gibbs-like state.

The paper is organized as follows. In Sec. \ref{model} we present
the finite-size closed Anderson model and outline the implementation
of the probes. In Section \ref{methods} we present our developed
numerical and analytic treatments of the density matrix dynamics:
First, we extend the standard quantum Langevin equation approach to
include reservoirs dynamics. The method can treat both the
noninteracting model (\ref{qle-u0}), the case with interactions,
only at the level of mean-field (Hartree) theory
(\ref{qle-ufinite}), and the probe model (\ref{qle-probe}). Second,
we present the fermionic trace formula, useful for studying the
Anderson model without interactions and with implemented dephasing
and inelastic effects in Sec.  \ref{levitov}. The third method,
presented in  Sec. \ref{infpi}, is an influence-functional path
integral approach \cite{dvira_prb,dvira_pccp}. This non-perturbative
tool can treat the model with interactions in a numerically exact
manner.
Applications are included in Sec. \ref{cd}. The effects of Coulomb
repulsion effects on the dot are studied using mean-field QLE and
the path integral technique in Sec. \ref{ee}. In Sec. \ref{dcd}
quantum equilibration and thermalization is investigated
using the probe technique. In this case, the total density matrix is
resolved using the QLE method and the fermionic trace formula. The
paper is summarized, along with an outlook, in Sec.
\ref{conclusions}.

\section{Model}
\label{model}

The closed-system Anderson model consists two metal grains,
$\nu=L,R$, including (each) a collection of $N_{\nu}$ dense
electronic levels initially populated by noninteracting electrons up
to the chemical potential $\mu_{\nu}$, at temperature
$T=\beta^{-1}$. The two baths couple only through their (weak)
hybridization with a single level quantum dot. Work presented in
this study concerns three variants of the model. The simplest
version is the ``noninteracting case", where electron-electron
repulsion effects and any decoherring and relaxation mechanisms are
excluded. The second case, the ``interacting model", allows for e-e
interactions on the dot only. The third model variant, the ``probe
model", phenomenologically contains elastic decoherring and
inelastic scattering processes on the dot, using the probe technique
and excluding e-e repulsion effects. This model is discussed in
detail in Sec. \ref{dcd}, where it is applied in the context of
quantum equilibration.

\subsection{The interacting model}

In the absence of decoherence and
dissipation effects the interacting Hamiltonian takes the form
\begin{equation}
H=H_{L}+H_{R}+H_{W}+\mathcal{V}_{L}+\mathcal{V}_{R},
\label{eq:mainH}
\end{equation}
where $H_{L,R,W}$ represents the Hamiltonian for the left reservoir,
right reservoir and the dot, respectively. The term $\mathcal{V}_{\nu}$
denotes the coupling of the dot to the $\nu$ reservoir,
\begin{eqnarray}
H_{L}&=&\sum_{l,\sigma}\epsilon_{l}c_{l,\sigma}^{\dagger}c_{l,\sigma},\,\,\,
H_{R}=\sum_{r,\sigma}\epsilon_{r}c_{r,\sigma}^{\dagger}c_{r,\sigma}
\nonumber\\
{\mathcal{V}_{L}}&=&\sum_{l\sigma}v_{l}c_{d,\sigma}^{\dagger}c_{l,\sigma}+h.c.
\,\,\,\, {\mathcal{V}_{R}}
=\sum_{r\sigma}v_{r}c_{d,\sigma}^{\dagger}c_{r,\sigma}+h.c.
\nonumber\\
H_{W}&=&\sum_{\sigma}\epsilon_{d}c_{d,\sigma}^{\dagger}c_{d,\sigma}
+ Un_{d,\uparrow} n_{d,\downarrow}.
\label{eq:H}
\end{eqnarray}
Here, $c_{k,\sigma}$ ($k=l,r,d$) are fermionic operators of the left
reservoir, $l\in L$, right reservoir, $r\in R$ and the dot ($d$).
The symbol $\sigma$ stands for the spin state ($\uparrow$ or
$\downarrow$) and $U$ accounts for the onsite repulsion energy. We
assume that $v_l$ and $v_r$ are real numbers and that the
Hamiltonians of the leads are diagonal in momentum basis and define
the hybridization $\Gamma_{\nu}(\epsilon)=\pi\sum_{k\in \nu}
v_k^2\delta(\epsilon-\epsilon_k)$, taken in practice to be
energy independent.
The Hamiltonian (\ref{eq:H}) disregards magnetic fields, yielding
spin-degenerate energy levels, thus it is sufficient to consider
observables for one spin species. We note that the noninteracting
case arises simply from the suppression of $U$.

Our objective in this paper is to calculate the time evolution of the expectation
values of {\it all} two-body operators in the system ($k,j=l,r,d$)
\bea \rho_{k,j}(t)\equiv \langle c_{k}^{\dagger}(t)c_j(t)\rangle
\equiv {\rm Tr}[\rho(t_0)c_k^{\dagger}(t)c_j(t)],
\label{eq:obj}
\eea
written here in the Heisenberg representation with
$\rho(t_0)=\rho_d\otimes \rho_L\otimes \rho_R$ representing the
factorized time-zero density matrix of the system, and with the
trace performed over all degrees of freedom. We suppress the spin
degree of freedom in the density matrix since its elements are
identical for the two spin configurations. As an initial condition,
we take the dot to be empty and the reservoirs' DM to be diagonal,
\bea \langle c_d^{\dagger}(t_0)c_d(t_0)\rangle =0,\,\,\, \langle
c_l^{\dagger}(t_0)c_l(t_0)\rangle =f_L(\epsilon_l)\equiv f_l, \,\,\,
\langle c_r^{\dagger}(t_0)c_r(t_0)\rangle&=&f_R(\epsilon_r)\equiv
f_r, \label{eq:init} \eea
with the population
$f_{L,R}(\epsilon)=[e^{\beta(\epsilon-\mu_{L,R})}+1]^{-1}$. As a
convention, we use the symmetric chemical potential bias
 $\mu_L=-\mu_R>0$.


\subsection{The probe model}

The Anderson probe model, a variant of the basic model, Eq.
(\ref{eq:H}), can emulate memory loss and energy redistribution in a
quantum system without explicitly introducing many-body interactions
\cite{Buttiker1,Buttiker2,Beenakker}. The probe technique has been
of extensive use in mesoscopic physics, for describing the
disappearance of quantum effects in transport \cite{Buttiker2},
dissipation \cite{Buttiker1}, and equilibration dynamics
\cite{Buttiker2b}. Recent advances include a full-counting
statistics analysis of the probe model \cite{Buttiker3}, and an
extension of the probe technique to the AC regime \cite{Buttiker4}.
We model here either a dephasing probe, allowing for quasi-elastic
decoherence processes, or a voltage probe, where inelastic effects
are further mimicked. In both cases we suppress electron-electron
interaction effects in the system.

As we explain next, in our study the ``probe" terminology refers to
a setup slightly different from the conventional one. The standard
construction refers to an open system scenario, where the probe
practically performs which-path experiments through repetitive
measurements of the system \cite{ExpVolt}. In contrast, in our
picture the probe is a finite-closed quantum system, only {\it
initialized} with a certain-special distribution. After its
preparation, the probe, similarly to other parts of the system, is
left undisturbed. Thus, we can use exact unitary approaches and
simulate the dynamics of the total system. While this picture abuses
to some extent the standard notion of a ``probe", we maintain this
terminology here since practically our implemented probe acts like a
proper one, inducing phase loss or/and energy reorganization in the
system.

We introduce a probe into the model by adding an additional
Fermi-sea reservoir, denoted by the letter $G$, to the Hamiltonian
(\ref{eq:H}), again discarding the spin degree of freedom,
\bea H_P=H+H_G+\mathcal{V}_G, \label{eq:Hprobe} \eea
where
\bea
H_G=\sum_{g}\epsilon_gc_g^{\dagger}c_g, \,\,\,\,
\mathcal{V}_G=\sum_g v_gc_{g}^{\dagger}c_d +h.c.
\eea
We naturally define the hybridization $\Gamma_G(\epsilon)=\pi\sum_g
v_g^2\delta(\epsilon-\epsilon_g)$, and take it as a constant. As
always, our objective here is the resolution of all system
expectation values of two-body operators ($k,j=l,r,d$),
$\rho_{k,j}(t)$. As initial conditions we assume Eq.
(\ref{eq:init}), where the $G$ bath initial condition is set
according to the particular probe condition, explained below.

{\it Voltage probe.} Inelastic scattering effects of electrons on
the dot are effectively included by implementation of a voltage
probe. The probe has a canonical distribution,
$f_G(\epsilon)=[e^{\beta(\epsilon-\mu_G)}+1]^{-1}$, and its chemical
potential $\mu_G$ is set such that the net charge current from the
dot to the $G$ unit vanishes for all times
\bea i_G\equiv \frac{d}{dt}\sum_g{\langle c_g^{\dagger}c_g\rangle
}=0.
\label{eq:ig}
\eea
With the motivation to explore situations beyond the linear response
regime \cite{DharSC}, we retrieve $\mu_G$ numerically, by employing
the Newton-Raphson method \cite{NR},
\bea \mu_G^{(m+1)}=\mu_G^{(m)}-i_G(\mu_G^{(m)})/i_G'(\mu_G^{(m)}).
\label{eq:NR} \eea
$\mu_G^{(0)}$ is the initial guess, $i'_G$ denotes the first
derivative with respect to $\mu_G$. In principle, one should adjust
$\mu_G$ throughout the simulation, to eliminate population leakage
from the $L$-dot-$R$ system into $G$. However, we have
found in our simulations that the $G$ bath has lawfully behaved as a
probe once we determined $\mu_G$ from the steady-state limit
using the following analytic expression for the charge current 
\bea
i_G(\epsilon)&=& \frac{2\Gamma_G}{\pi} \frac{\Gamma_R
[f_G(\epsilon)-f_R(\epsilon)] + \Gamma_L
[f_G(\epsilon)-f_L(\epsilon)] } {(\epsilon-\epsilon_d)^2+\Gamma^2},
\nonumber\\
i_G&=& \int  i_G(\epsilon) d\epsilon,
\label{eq:igA}
\eea
with $\Gamma=\Gamma_{L}+\Gamma_R+\Gamma_G$ \cite{comm}.
The lower and upper integration limits are determined by the band simulated.
Substituting
Eq. (\ref{eq:igA}) in Eq. (\ref{eq:ig}), a voltage probe condition
is set by demanding $f_G(\epsilon)$ to fulfill the relation
\bea \int d\epsilon
\frac{f_G(\epsilon)}{(\epsilon-\epsilon_d)^2+\Gamma^2} =
\frac{1}{\Gamma_L+\Gamma_R} \int d\epsilon
\frac{f_L(\epsilon)\Gamma_L + f_R(\epsilon)\Gamma_R}
{(\epsilon-\epsilon_d)^2+\Gamma^2}. \label{eq:voltage} \eea
{\it Dephasing probe.} Implementation of the dephasing probe,
fabricating elastic decoherence, necessitates the stronger
requirement $i_G(\epsilon)=0$, i.e., the charge current at a given
energy should vanish. Using the steady-state behavior
(\ref{eq:igA}), we obtain a non-Fermi distribution
\bea
f_{G}(\epsilon)=\frac{\Gamma_{R}f_{R}(\epsilon)+\Gamma_{L}f_{L}(\epsilon)}{\Gamma_{R}+\Gamma_{L}}.
\label{eq:dephasing} \eea
We emphasize that $\mu_G$ or $f_G$ have been determined here in the
steady-state limit, assuming fixed chemical potentials for the $L$
and $R$ baths. Indeed, at short time, $\Gamma_{L,R} t\lesssim2$,
before a (quasi) steady-state sets in, we find that $i_G\neq0$.
However, we have confirmed numerically that beyond this time
throughout all our simulations $|i_G/i_{L,R}|<10^{-4}$, thus the $G$
reservoir plays the role of a proper probe.

Three different approaches for the calculation of the full DM are
described in Sections \ref{sec-qle}, \ref{levitov} and \ref{infpi}.
Applications are included in Sec. \ref{cd}.


\section{Methods}
\label{methods}

\subsection{Quantum Langevin Equation}
\label{sec-qle}

The dynamics of the Anderson model in the absence of interactions
($U=0$), with interactions at the mean-field level, or with a probe,
is described here within a quantum Langevin equation framework
\cite{ford}. The basis of our method has been used in the past to
follow the dot evolution or the charge and energy currents in the
system \cite{sen,DharDM}. Here, we show results for the full DM. We begin
our analysis with the trivial treatment of the impurity (dot) and review
the steps involved. This review helps highlight underling approximations and establishes
limits for the method's applicability.

\subsubsection{Noninteracting case ($U=0$)}
\label{qle-u0}

In the Heisenberg representation the fermionic operators satisfy the following
equations of motion (EOM),
\begin{eqnarray}
\dot{c}_{d}&=&-i\epsilon_{d}c_{d}-i\sum_{l}v_{l}c_{l}-i\sum_{r}v_{r}c_{r}
\label{eq:EOMd}
\nonumber\\
\dot{c}_{l} &=&-i\epsilon_{l}c_{l}-iv_{l}c_{d}
\label{eq:EOML}
\nonumber\\
\dot{c}_{r}&=&-i\epsilon_{r}c_{r}-iv_{r}c_{d}
\label{eq:EOM}
\end{eqnarray}
Formal integration of the  reservoirs EOM yields, e.g. at the $L$ end,
\begin{eqnarray}
c_{l}(t)&=&e^{-i\epsilon_{l}(t-t_{0})}c_{l}(t_{0})-
iv_{l}\int_{t_{0}}^{t}d\tau
e^{-i\epsilon_{l}(t-\tau)}c_{d}(\tau)d\tau.  \label{eq:clm}
\end{eqnarray}
We substitute Eq. (\ref{eq:clm}), and the analogous expression for $c_r(t)$,
into the dot EOM [Eq. (\ref{eq:EOM})], and retrieve
\begin{eqnarray}
\dot{c}_{d}&=&-i\epsilon_{d}c_{d} -
i\sum_{l}v_{l}e^{-i\epsilon_{l}(t-t_{0})}c_{l}(t_{0})
-i\sum_{r}v_{r}e^{-i\epsilon_{r}(t-t_{0})}c_{r}(t_{0})
\nonumber\\
&-&\int_{t_{0}}^{t}d\tau\sum_{l}v_{l}^{2}
e^{-i\epsilon_{l}(t-\tau)}c_{d}(\tau)
-\int_{t_{0}}^{t}d\tau\sum_{r}v_{r}^{2}e^{-i\epsilon_{r}(t-\tau)}c_{d}(\tau).
 \label{eq:qle}
\end{eqnarray}
In this (exact) equation the second and third terms are interpreted as
 ``noise'' \cite{ford},
\begin{eqnarray}
\eta^{L}(t)&\equiv&\sum_{l}v_{l}e^{-i\epsilon_{l}(t-t_{0})}c_{l}(t_{0})
\nonumber\\
\eta^{R}(t) &\equiv&\sum_{r}v_{r}e^{-i\epsilon_{r}(t-t_{0})}c_{r}(t_{0}).
\label{eq:noise}
\end{eqnarray}
The last two terms in Eq. (\ref{eq:qle}) can be reduced, each, into
decay terms, further inducing an energy shift of the dot energy,
absorbed into the definition of $\epsilon_d$. This is justified by
following two assumptions: (i) The hybridization
$\Gamma_{\nu}(\epsilon)=\pi\sum_{k\in\nu} v_k^2
\delta(\epsilon-\epsilon_k)$ may be taken as a positive real
self-energy function \cite{ford}, and (ii) the dot dynamics is slow
relative to the reservoirs' evolution. We now explain the steps
involved. First, we define a new operator for the dot, by absorbing
its fast oscillatory behavior, $\tilde c_d(t)\equiv
c_de^{i\epsilon_dt}$. Its EOM is
\bea
\dot{\tilde c}_{d}&=&
-i[\eta_L(t)+\eta_R(t)] e^{i\epsilon_dt}
-\int_{t_{0}}^{t}d\tau\sum_{l}v_{l}^{2}
e^{-i\epsilon_{ld}(t-\tau)}\tilde c_{d}(\tau)
-\int_{t_{0}}^{t}d\tau\sum_{r}v_{r}^{2}e^{-i\epsilon_{rd}(t-\tau)}\tilde
c_{d}(\tau) \label{eq:A2} \eea
where $\epsilon_{kj}\equiv\epsilon_k-\epsilon_j$. We then change
variables, $x\equiv t-\tau$, and make the assumption that the dot
evolution (now missing the fast phase oscillation) is slow with
respect to other time scales in the system, $\tilde c_d(t-x)\sim
\tilde c_d(t)$. This results in
\bea \int_{t_{0}}^{t}d\tau\sum_{l}v_{l}^{2}
e^{-i\epsilon_{ld}(t-\tau)}\tilde{c}_{d}(\tau)\approx\tilde{c}_{d}(t)\sum_{l}v_{l}^{2}\int_{0}^{t-t_{0}}e^{-i\epsilon_{ld}x}dx
\nonumber \eea
If the time is long, $t\gg t_0$, integration gives
\bea
&&
\sum_{l}v_{l}^{2}\int_{0}^{t\rightarrow
\infty}e^{-i\epsilon_{ld}x}dx = \Gamma_{L}(\epsilon_{d})
-2iv_{l}^{2}
\lim_{t\rightarrow\infty}\left[\int_{-\infty}^{+\infty}D_L(\epsilon)\frac{\sin^{2}(\epsilon-\epsilon_{d})t}{\epsilon-\epsilon_{d}}d\epsilon\right]
\nonumber \eea
with $D_L(\epsilon)=\sum_{l}\delta(\epsilon-\epsilon_l)$ as the
density of states of the $L$ metal, taken as flat here.
We also take the interaction parameters $v_l$ to be independent 
of the $l$ index.
The imaginary term introduces an
energy shift, which can be absorbed into the definition of
$\epsilon_d$. It diminishes when the density of states does not
depend on energy (the case used later), and when the bandwidth is
large enough. In our numerical calculations we have used a finite
bandwidth with a cutoff  $D=\pm 1$, introducing a small correction
to $\epsilon_d$.
We return to Eq. (\ref{eq:qle})  and conclude that it obeys the
quantum Langevin equation
\bea \dot c_d =-i\epsilon_d c_d -i\eta^L(t) -i\eta^R(t)
-\Gamma(\epsilon_d) c_d(t), \label{eq:cdQLE} \eea
with $\Gamma(\epsilon)=\Gamma_L(\epsilon)+\Gamma_R(\epsilon)$. The
dynamics of the dot occupation, $\langle n_d(t) \rangle$, can be
reached by a formal integration of Eq. (\ref{eq:cdQLE}),
\bea c_d(t)=c_d(t_0)e^{(-i\epsilon_d-\Gamma)(t-t_0)}
-i\int_{t_0}^{t}e^{(-i\epsilon_d-\Gamma)(t-\tau)}
[\eta^L(\tau)+\eta^R(\tau)]d\tau, \label{eq:cdi} \eea
to provide the standard expression \cite{Komnik}
\bea \langle n_{d}(t)\rangle &\equiv&\langle c_d^{\dagger}(t)c_d(t)
\rangle=\sum_{k=l,r}\frac{|v_{k}|^{2}f_{k}}{\epsilon_{dk}^{2}+\Gamma^{2}}
\Big[1+e^{-2\Gamma(t-t_{0})}
-2e^{-\Gamma(t-t_{0})}\cos[\epsilon_{dk}(t-t_{0})]\Big].
\label{eq:dot-occ}
\eea
This derivation relied on the initial conditions (\ref{eq:init}).
The summation runs over all the reservoirs degrees of freedom
\cite{foot_dense}. We now use Eq. (\ref{eq:clm}) and its analogous
expression for $c_r(t)$, together with Eq. (\ref{eq:cdi}), and
derive analytical expressions for other quadratic expectation
values, $\langle c_k^{\dagger}(t)c_j(t)\rangle$, $k,j=l,r,d$. These
results are valid as long as one can faithfully rely on Eq.
(\ref{eq:cdQLE}). In what follows we take $t_0=0$ to simplify our
notation. The reservoir-dot coherence can be obtained analytically,
%
\bea \rho_{d,l}(t)\equiv \left\langle
c_{d}^{\dagger}(t)c_{l}(t)\right\rangle =B_{1}+B_{2}.
\label{eq:deco_dot_res} \eea
Here, $B_1$ includes contributions from the $L$ side only,
\bea
B_{1}=\frac{iv_{l}f_{l}}{\Gamma-i\epsilon_{dl}}\left[1-e^{-t(\Gamma-i\epsilon_{dl})}\right].
\eea
$B_2$ includes electron transmission pathways from the $L$ side,
through the dot, to the $K=L,R$ grain,
\bea &&B_{2}= -iv_{l}\sum_{k\in
L,R}\frac{v_{k}^{2}f_{k}}{\Gamma^{2}+\epsilon_{dk}^{2}} \nonumber\\
&&\times \Biggl\{\frac{1-e^{i\epsilon_{kl}t}}{i\epsilon_{lk}} +
\frac{e^{-2\Gamma
t}-e^{-t(\Gamma-i\epsilon_{dl})}}{i\epsilon_{ld}-\Gamma}
-\frac{e^{-t(\Gamma-i\epsilon_{kd})}-
e^{-i\epsilon_{lk}t}}{i\epsilon_{ld}-\Gamma}
-\frac{e^{-t(\Gamma-i\epsilon_{dk})}-e^{-t(\Gamma+i\epsilon_{ld})}}
{i\epsilon_{lk}}\Biggr\}.
\eea
Using Eq. (\ref{eq:deco_dot_res}), we derive an
expression for the charge current at the $L$ contact,
\bea
 i_{L}(t)&\equiv&
\frac{d}{d t}
\sum_l \langle c_l^{\dagger}(t)c_l(t)\rangle=
-2\Im\sum_{l}v_{l}\left\langle
c_{d}^{\dagger}(t)c_{l}(t)\right\rangle
\nonumber\\
& =&
\frac{2\Gamma_{L}\Gamma_{R}}{\pi}\sum_{l}\frac{f_{L}-f_{R}}{\epsilon_{ld}{}^{2}+\Gamma^{2}}
-\frac{2\Gamma_{L}}{\pi}e^{-\Gamma t}
\sum_{l}\frac{1}{\epsilon_{ld}{}^{2}+\Gamma^{2}}
\nonumber\\
&\times& \Biggl\{ e^{-\Gamma t}(f_{l}\Gamma_{L}+f_{r}\Gamma_{R})
-(2f_{l}\Gamma_{L}+2f_{r}\Gamma_{R}-\Gamma
f_{l})\cos\left(\epsilon_{ld}t\right) -f_{l}
\epsilon_{ld}\sin\left(\epsilon_{ld}t\right)\Biggr\}.
\label{eq:im_der} \eea
Here $\Im$ stands for the imaginary part.
An analogous expression can be written for $i_R(t)$. Ref.
\cite{Komnik} includes a Green's function based derivation for the
time dependent current in the symmetric limit ($\Gamma_L=\Gamma_R$).
This derivation results in a surplus nonphysical term at the initial
time. We now turn our attention to the reservoirs' states
population. Using Eq. (\ref{eq:clm}), we find that it is given by
three contributions,
\bea
p(\epsilon_l)&\equiv& \langle c_l^{\dagger}(t)c_l(t)\rangle =\langle
c_l^{\dagger}(t_0)c_l(t_0)\rangle
\nonumber\\
&+&iv_le^{-i\epsilon_l(t-t_0)} \int_{t_0}^{t}
e^{i\epsilon_l(t-\tau)} \langle c_d^{\dagger}(\tau)c_l(t_0) \rangle
d\tau + c.c.
\nonumber\\
&+& v_l^2 \int_{t_0}^{t}\int_{t_0}^td\tau_1 d\tau_2 \langle
c_{d}^{\dagger}(\tau_1)c_d(\tau_2)\rangle
e^{i\epsilon_l(t-\tau_1)}e^{-i\epsilon_l(t-\tau_2)}. \label{eq:pl}
\eea
The two-times correlation functions can be obtained from
Eqs. (\ref{eq:clm}) and (\ref{eq:cdi}) without additional
approximations, and the explicit expressions
are given in Appendix A. Similarly, closed analytic expressions can
be written for inter and intra-reservoir coherences, e.g.
$\rho_{l,k}(t)$, $k\in L, R$, see Appendix A. Eqs.
(\ref{eq:dot-occ}), (\ref{eq:deco_dot_res}), (\ref{eq:pl}),
(\ref{eq:non-dia}) and (\ref{eq:diag-like}), and the analogous
$R$-bath expressions, form the time-dependent full density matrix of
the system.

{\it Timescales.} We now comment on the applicability of the QLE
approach. Given infinite reservoirs, a current carrying steady-state
behavior develops, and the dot occupation, as well as the charge
current, reach a fixed value after a short time, $\tau_t
\gtrsim2/\Gamma$ (see for example Fig. \ref{FigINFPI5}). However,
since the reservoirs are finite in the present treatment, recurrence
effects should eventually manifest in our system. These effects
cannot be handled by the QLE technique since an {\it irreversible}
behavior has been assumed for the dot, as Eq. (\ref{eq:cdQLE})
breaks unitary evolution. The technique can still excellently
reproduce the exact dynamics in the so called ``quasi steady-state"
(QSS) region, up to $\tau_d\sim 2\pi N/D$ \cite{yuka}.
Here $N$ is the number of electronic states in
each bath and $D$ is the band cutoff. Within this time, the dot
occupation and the charge current are constant, similar to a real
steady-state situation. Around the time $\tau_d$ the dot occupation
should begin to vary, showing (partial) recurrence behavior,
and the QSS limit breaks down.

\begin{figure}[htbp]
\vspace{0mm} \hspace{0mm} {\hbox{\epsfxsize=90mm \epsffile{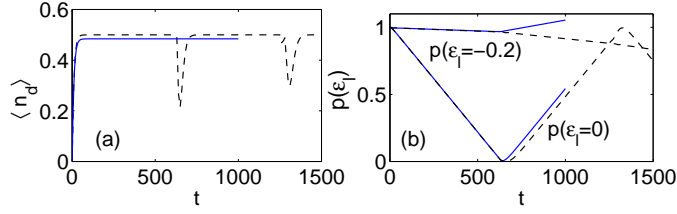}}}
\caption{ (a) Dot occupation as a function of time
calculated using either the QLE method (full line)
or the trace formula (dashed line), described in Sec. \ref{levitov}.
(b) Population of two selected states of the left reservoir, plotted
 as a function of time, $\epsilon_l=-0.2$ (top) and $\epsilon_l=0$ (bottom).
In both cases we simulated the noninteracting Anderson model with
$\epsilon_d=0$, $\mu_L=-\mu_R=0.2$, $\Gamma_L=\Gamma_R=0.025$
and $\Gamma_G=0$, $N_{L,R}$=201 and $D=\pm1$.}
\label{Figtau}
\end{figure}

Fig. \ref{Figtau} clarifies this timescale issue. The left panel
displays the dot occupation as a function of time using either the
QLE approach (full line) or an exact method (dashed line), described
in Sec. \ref{levitov}. Results agree up to $t\sim \tau_d\sim630$,
and deviations before this time are due to the finite band used in
QLE, while neglecting the energy correction in Eq. (\ref{eq:cdQLE}).
Around the time $\tau_d$ exact simulations show a partial reversal
of the dot occupation, while QLE still produces the QSS value. At a
later time, $t\gg \tau_d$, the QLE data diverges. Panel (b) presents
the bath occupation for two selected energies, and we find that
nonphysical values, such as a population exceeding unity, can be
obtained with QLE when $t>\tau_d$. Thus, the QLE method can be used
within the interval $t<\tau_{d}$ only, to be consistent with its
underlying assumptions. However, interesting nonequilibrium physics
takes place within this window, thereby making this
approach valuable considering that exact computational schemes are
very expensive.

\subsubsection{Mean-field theory}
\label{qle-ufinite}

The QLE description of Sec. \ref{qle-u0} can be generalized to
accommodate electron-electron repulsion effects on the dot at the
mean-field level. We refer to this extension as a MF QLE treatment, and note
that it is not trivial: While a MF theory has been developed, suffering from
some pathologies, for the study of dot occupation or charge current in the
steady-state limit \cite{KomnikMF,BerMF}, here we present a MF scheme
to describe the {\it real-time dynamics} of the {\it full
density matrix}. By comparing MF results to exact numerical
simulations, see Sec. \ref{cd}, we conclude that a MF description
can produce physical results up to
$\frac{U}{\Gamma},\frac{U}{\mu_{L}-\mu_{R}}\lesssim O(1)$. The
effectiveness of the method also delicately depends on the dot level
position, see for example Fig. \ref{FigINFPI3}.

The MF prescription, treating Coulombic repulsion, takes us
back to Eq. (\ref{eq:H}). We now assume that the many-body interaction term can be factorized
\cite{KomnikMF,BerMF}
\bea Un_{d,\uparrow}n_{d,\downarrow}\rightarrow U\left[\left\langle
n_{d,\uparrow}\right\rangle
n_{d,\downarrow}+n_{d,\uparrow}\left\langle
n_{d,\downarrow}\right\rangle \right]. \label{MFprescription} \eea
This assumption reduces the Hamiltonian
to an effectively noninteracting one, with a renormalized dot energy
\bea \tilde{\epsilon}_{d}(t)=\epsilon_{d}+U\left\langle
n_{d}(t)\right\rangle. \label{eq:edtilde} \eea
The spin-index has been dropped here, as we choose not to study magnetic effects.
The formalism could be feasibly generalized to include
magnetic fields, resulting in
$\epsilon_{d,\uparrow}\neq\epsilon_{d,\downarrow}$. In such
situations the validity of MF equations is governed by another
energy scale besides $U/\Gamma$ and $U/\Delta \mu$, namely
$U/(\epsilon_{d,\uparrow}-\epsilon_{d,\downarrow})$. The dot
occupation  is  determined in a self-consistent manner at every
instant by modifying Eq. (\ref{eq:dot-occ}) to contain the dot
renormalized energy,
\bea \langle n_{d}(t)\rangle =
\sum_{k=l,r}\frac{|v_{k}|^{2}f_{k}}{\left[\epsilon_{dk}+U \langle
n_{d}(t)\rangle\right]^{2} +\Gamma^{2}}  \Big[1+e^{-2\Gamma t}
-2e^{-\Gamma t}\cos[(\epsilon_{dk}+U \langle
n_{d}(t)\rangle)t]\Big]. \label{eq:MFdot-occ} \eea
The solution provides the renormalized dot energy
$\tilde{\epsilon}_{d}(t)$, which is then used to replace
$\epsilon_d$ in Eqs. (\ref{eq:deco_dot_res}), (\ref{eq:pl}),
(\ref{eq:non-dia}) and (\ref{eq:diag-like}), to provide the full DM
at the MF level.

\subsubsection{Probe model}
\label{qle-probe}

To implement elastic dephasing or inelastic effects with a probe,
the set of equations (\ref{eq:EOM}) is augmented by an additional
equation for $c_g$. The EOM for $c_d$ must be modified to include
its coupling to the $G$ bath,
\bea \dot{c}_{g} &=&-i\epsilon_{g}c_{g}-iv_{g}c_{d}
\nonumber\\
\dot{c}_{d}&=&-i\epsilon_{d}c_{d}-i\sum_{l}v_{l}c_{l}-i\sum_{r}v_{r}c_{r}
-i\sum_g v_gc_g. \eea
It can be easily shown that under the QLE basic assumptions, as
discussed in Sec. \ref{qle-u0}, the dot still satisfies Eq.
(\ref{eq:cdQLE}) with an additional noise term $\eta^G$ and with a
re-defined total hybridization, $\Gamma=\Gamma_L+\Gamma_R+\Gamma_G$.
The noise $\eta^G$ obeys a relation analogous to Eq. (\ref{eq:noise}),
and Eq. (\ref{eq:cdi}) is generalized to
\bea c_d(t)=c_d(t_0)e^{(-i\epsilon_d-\Gamma)(t-t_0)}
-i\int_{t_0}^{t}e^{(-i\epsilon_d-\Gamma)(t-\tau)}
[\eta^L(\tau)+\eta^R(\tau)+ \eta^G(\tau)]d\tau.
\label{eq:cdin} \eea
We can now recognize that, in the presence of the probe, the expressions for
the DM elements (\ref{eq:dot-occ}), (\ref{eq:deco_dot_res}),
(\ref{eq:pl}), (\ref{eq:non-dia}) and (\ref{eq:diag-like}) stay
formally intact. The technical adjustments are as follows: (i) We
re-define the total hybridization,
$\Gamma=\Gamma_L+\Gamma_R+\Gamma_G$. (ii) We augment summations that
run over both $L$ and $R$ baths by $k\in G$ terms. For example, the
summation in $F_3$ [Eq. (\ref{eq:F2F3})] should include such terms.
(iii) We set the $G$ bath distribution to satisfy the probe
conditions, explained in Sec. \ref{model}.

\subsection{Fermionic trace formula ($U=0$)}
\label{levitov}

We describe here an exact brute force calculation that can provide
numerically all the elements of the density matrix in the
noninteracting case. We begin without the presence of a probe, opting to include its effects
later. This unitary method complements
the QLE description, whose validity is governed by $\tau_{d}$.
Since the method is unitary, a recurrences behavior is expected to manifest
at long enough time.
The core of the method is the trace formula for fermions
\cite{israel}
\bea {\rm Tr}\left[e^{M_1}
e^{M_2}...e^{M_p}\right]=\det\left[1+e^{m_1}e^{m_2}...e^{m_p}\right],
\label{eq:trace} \eea
where $m_p$ is a single-particle operator corresponding to a
quadratic operator $M_p=\sum_{i,j}(m_p)_{i,j}c_i^{\dagger}c_j$.
$c_{i}^{\dagger}$ ($c_j$) are fermionic creation (annihilation)
operators. Our objective is the dynamics of a quadratic operator
$A$, either given by system or bath degrees of freedom,
$A\equiv c_j^{\dagger}c_{k}$, $j,k=l,r,d$,
\bea \langle A(t) \rangle = {\rm Tr}\left[\rho(t_0)e^{iHt}Ae^{-iHt}\right] =
{\rm lim}_{\lambda \rightarrow 0}\frac{\partial}{\partial \lambda}
{\rm Tr} \left[\rho_L\rho_R\rho_d e^{iHt} e^{\lambda A}
e^{-iHt}\right]. \label{eq:exact} \eea
We introduce the $\lambda$ parameter, taken to vanish at the end of
the calculation. The initial condition is taken to be factorized,
$\rho(t_0)=\rho_d\otimes\rho_L\otimes \rho_R$,
$\rho_{\nu}=e^{-\beta(H_{\nu}-\mu_{\nu}N_{\nu})}/Z_{\nu}$, $Z_{\nu}$
is the partition function, $\rho_d$ describes the dot initial
density matrix. These density operators follow an exponential form,
$e^{M}$, with $M$ a quadratic operator. The application of the trace
formula leads to
\bea \langle e^{\lambda A(t)}
\rangle=\det\left\{[I_L-f_L]\otimes[I_R-f_R]\otimes[I_d-f_d] +
 e^{iht}e^{\lambda
 a}e^{-iht} f_L\otimes f_R \otimes f_d\right\}
\label{eq:trace2}.
\eea
Here, $a$ and $h$ are single-body matrices of the $A$ and $H$
operators, respectively. The matrices $I_{\nu}$ and $I_d$ are the
identity matrices for the $\nu=L,R$ space and for the dot. The
functions $f_L$ and $f_R$ are the band electrons occupancy
$f_{\nu}(\epsilon)=[e^{\beta(\epsilon-\mu_{\nu})}+1]^{-1}$. Here
they are written in matrix form and in the energy representation.
$f_d$ represents the dot initial occupation, again written in a
matrix form. Since we are working with finite-size reservoirs, Eq.
(\ref{eq:trace2}) can be readily simulated numerically-exactly.

The fermionic trace formula can be trivially generalized to include
a probe. We add the $G$ bath into the expectation value expression,
$A\equiv c_j^{\dagger}c_{k}$,
\bea \langle A(t) \rangle = {\rm
Tr}\left[\rho(t_0)e^{iH_Pt}Ae^{-iH_Pt}\right] = {\rm lim}_{\lambda
\rightarrow 0}\frac{\partial}{\partial \lambda} {\rm Tr}
\left[\rho_L\rho_R\rho_G\rho_d e^{iH_Pt} e^{\lambda A}
e^{-iH_Pt}\right], \label{eq:exactG} \eea
where as before
$\rho_{\nu}=e^{-\beta(H_{\nu}-\mu_{\nu}N_{\nu})}/Z_{\nu}$, $Z_{\nu}$
is the partition function, $\nu=L,R,G$. $\rho_d$ stands for the dot
initial density matrix, and we trace over all DOF, the two
reservoirs, the probe, and the dot.

{\it Timescales.} Simulations with the trace formula are not
restricted to a certain time scale. The method is unitary, providing
(physical) recurrence behavior due to finite size effects.
Since the time evolution scheme is not iterative, the
accuracy of results does not deteriorate in time.

\subsection{Numerically exact path integral simulations,  $U \neq 0$}
\label{infpi}

The time evolution of the closed and {\it interacting} Anderson
model can be simulated by employing a numerically-exact iterative
influence-functional path integral (INFPI) approach
\cite{dvira_prb,dvira_pccp}. This method relies on the fact that in
out-of-equilibrium (and nonzero temperature) cases bath correlations
have a finite range, allowing for their truncation beyond a memory
time dictated by the voltage-bias and the temperature. Based on this
finite-memory assumption, an iterative-deterministic time-evolution
scheme has been developed, where convergence with respect to the
memory length can, in principle, be reached. The principles of the
INFPI approach have been detailed in Refs.
\cite{dvira_prb,dvira_pccp}, where it has been developed to
investigate dissipation effects in the nonequilibrium spin-fermion
model, and the population and current dynamics in correlated
quantum dots. Recently, it has been used to examine the effects of
a magnetic flux on the intrinsic coherence dynamics in an
Aharonov-Bohm quantum dot interferometer \cite{AB}.
The INFPI method relies on the existence of a finite decorrelation
time, thus it is suited for simulating the dynamics of an impurity
coupled to a bath. Here we show that it can be used to
retrieve the total DM in a system-bath setup. While in principle
the method could encompass both interactions and probe, we focus exclusively on
the first element.

The method is based on the fermionic trace formula
(\ref{eq:trace}),  incorporating many-body effects within a
path integral expression. Our work starts with the time evolution
expression (\ref{eq:exact}) under the Hamiltonian (\ref{eq:H}). We
factorize the time evolution operator, $e^{iHt} = (e^{iH\delta
t})^{N_t}$, $N_t\delta t=t$, and adopt the Trotter decomposition
$e^{iH\delta t}\approx\big( e^{iH_0\delta t/2} e^{iH_1 \delta t}
e^{iH_0\delta t/2} \big)$, where $H=H_0+H_1$ with
\bea H_0&=&\sum_{\nu=L,R}\left(H_{\nu}+\mathcal{V}_{\nu} \right)+
\sum_{\sigma}\left(
\epsilon_d+\frac{U}{2}\right)c_{d,\sigma}^{\dagger}c_{d,\sigma}
\nonumber\\
H_1&=& U\big[n_{d,\uparrow}n_{d,\downarrow} -\frac{1}{2} (n_{d,\uparrow} +n_{d,\downarrow})\big].
\eea
$H_1$ extracts many-body interactions on the dot, and it is eliminated
by introducing auxiliary Ising variables $s=\pm$ via the
Hubbard-Stratonovich (HS) transformation \cite{Hubb-Strat},
\bea e^{\pm iH_1 \delta t} = \frac{1}{2} \sum_{s} e^{H_\pm(s)}, \,\,\,\,\,\,\
e^{H_\pm(s)}\equiv e^{-s \kappa_{\pm}
(n_{d,\uparrow}-n_{d,\downarrow})}. \label{eq:HS} \eea
Here, $\kappa_{\pm}=\kappa' \mp i \kappa'' $,
$\kappa'=\sinh^{-1}[\sin(\delta t U/2)]^{1/2}$, $\kappa''
=\sin^{-1}[\sin (\delta t U/2)]^{1/2}$. The uniqueness of this
transformation requires that $U \delta t < \pi$. Incorporating the
Trotter decomposition and the HS transformation into Eq.
(\ref{eq:exact}), we find that the time evolution of $A$ is
dictated by
\bea \langle A(t)\rangle = \lim_{\lambda \rightarrow 0}
\frac{\partial}{\partial \lambda} \Big \{ \int ds_1^{\pm}
ds_2^{\pm}... ds_{N_t}^{\pm} I(s_{1}^{\pm}, s_2^{\pm},...,
s_{N_t}^{\pm}) \Big \}. \label{eq:At2} \eea
The integrand, referred to as as the ``Influence Functional" (IF), is given by
($q=1$, $q+p=N_t$)
\bea
 I(s_q^{\pm},..., s_{q+p}^{\pm})=
\frac{1}{2^{2(p+1) }}
{\rm Tr} \Big[\rho(t_0)
\mathcal
G_+(s_{q+p}^+) ... \mathcal G_+(s_{q}^+) e^{i H_0 (q-1) \delta
t} e^{\lambda  {A}} e^{-i H_0 (q-1) \delta t} \mathcal
G_-(s_{q}^-)... \mathcal G_-(s_{q+p}^-)  \Big],
\nonumber\\
\label{eq:IF}
\eea
where $\mathcal G_{+}(s_q^{+}) = \left( e^{i H_0 \delta t/2} e^{
H_{+}(s_{q}^{+})}  e^{i H_0 \delta t/2} \right)$ and $\mathcal
G_-=\mathcal G_+^{\dagger}$. Eq. (\ref{eq:At2}) is exact in the
$\delta t\rightarrow 0$ limit. Practically, it is evaluated by
truncating the IF beyond a memory time $\tau_c=N_s\delta t$,
corresponding to the time beyond which bath correlations may be
ignored \cite{dvira_prb}, $N_s$ is an integer.
The following (non-unique) breakup has been suggested by
\cite{dvira_prb},
\bea I(s_{1}^{\pm}, s_{2}^{\pm},...s_{N_t}^{\pm} ) \simeq
I(s_1^{\pm},s_2^{\pm},..., s_{N_s}^{\pm})
I_s(s_2^{\pm},s_3^{\pm},..., s_{N_s+1}^{\pm}) ...
I_s(s_{N_t-N_s+1}^{\pm},s_{N_t-N_s+2}^{\pm},..., s_{N_t}^{\pm}),
\label{eq:dynamics} \eea
where each element in the product, besides the first one, is given
by a ratio between truncated IFs,
\bea I_s(s_q,s_{q+1},...,s_{q+N_s-1})=
\frac{I(s_q^{\pm},s_{q+1}^{\pm},...,s_{q+N_s-1}^{\pm})}{I(s_{q}^{\pm},s_{q+1}^{\pm},...,s_{q+N_s-2}^{\pm})}.
 \label{eq:Is} \eea
We now define a multi-time object,
\bea &&{\mathcal R}(s_{q+1}^{\pm}, s_{q+2}^{\pm},...,
s_{q+N_s-1}^{\pm} )
\nonumber\\
&&\equiv \sum_{s_1^{\pm},s_2^{\pm},..., s_{q}^{\pm}}
I(s_1^{\pm},s_2^{\pm},..., s_{N_s}^{\pm})
I_s(s_2^{\pm},s_3^{\pm},..., s_{N_s+1}^{\pm})... \times
I_s(s_{q}^{\pm},s_{q+1}^{\pm},..., s_{q+N_s-1}^{\pm}), \label{eq:R1}
\eea
and evolve it iteratively by multiplication  with the subsequent
truncated IF, followed by summation over the time variables at the head,
\bea {\mathcal R}(s_{q+2}^{\pm}, s_{q+3}^{\pm},..., s_{q+N_s}^{\pm}
)= \sum_{s_{q+1}^{\pm}} {\mathcal R}(s_{q+1}^{\pm},
s_{q+2}^{\pm},..., s_{q+N_s-1}^{\pm} )
I_s(s_{q+1}^{\pm},s_{q+2}^{\pm},...,s_{q+N_s}^{\pm}). \label{eq:R2}
\eea
The behavior at a particular time $t_q$ is reached by summation over the
internal variables,
\bea \langle e^{\lambda A(t_q)} \rangle =
\sum_{s_{q+2-N_s}^{\pm},...,s_{q}^{\pm}} {\mathcal
R}(s_{q+2-N_s}^{\pm}, s_{q+3-N_s}^{\pm},..., s_{q}^{\pm} ).
\label{eq:R3} \eea
This procedure is repeated for several (small) values of
$\lambda$, and the expectation value $\langle A(t_q)\rangle$ is
retrieved by numerical differentiation in $\lambda$.
The truncated IF, Eq. (\ref{eq:IF}), is the core of this
calculation. It is achieved numerically-exactly
using the fermionic trace formula (\ref{eq:trace}).

{\it Timescale.} Previous studies for dense reservoirs have
confirmed that INFPI can provide accurate results in both short time
and in the quasi steady-state region \cite{dvira_prb,dvira_pccp}. However, the
method is not restricted to such dense-reservoirs situations, and it
can describe the dynamics of small metallic grains since it handles
all states explicitly. It should be still noted that the basic
working assumption behind INFPI is the existence of a finite bath-induced
decorrelation time. If the metal grains are very small,
including few discrete states, this memory time $\tau_c$ does not
exist or it becomes large, hindering convergence. Roughly, one could
expect that a decorrelation time can be identified when a
system-bath picture still holds, in the sense that a QLE description
can be written i.e., Eq. (\ref{eq:cdQLE}) is valid. In such
situations, INFPI simulations should converge and generally hold
beyond $\tau_d$. In practice, since these calculations are
intensive, we have computed dynamics within a relatively short
interval, $\Gamma t<5$, where the QSS description is still valid.


\begin{figure}[htbp]
\vspace{0mm} \hspace{0mm} {\hbox{\epsfxsize=80mm \epsffile{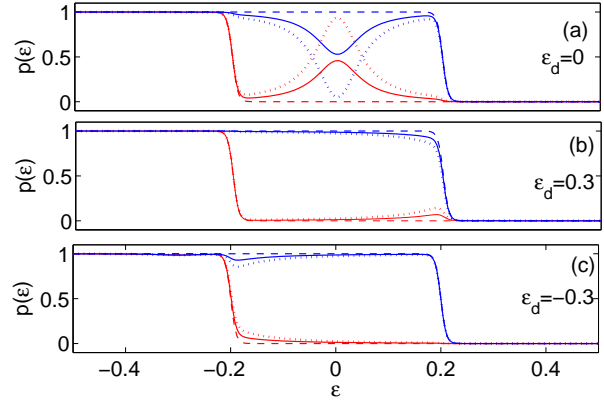}}}
\caption{Population of reservoirs' levels in the noninteracting case
with the dot energy positioned (a) within the bias window at
$\epsilon_d=0$, (b) above the bias window at $\epsilon_d=0.3$, and
(c) below it at $\epsilon_d=-0.3$. Plotted are the $L$ (three top
lines) and $R$ (three bottom lines) occupations as a function of
electron energy, at $\Gamma t=0$ (dashed) $\Gamma t=9.5$ (full) and
$\Gamma t=19$ (dotted). The framework used is a quantum Langevin
approach (Sec. \ref{sec-qle}) with $\beta=200$ for the inverse
temperature, $\Gamma_L=\Gamma_R=0.025$ for bath-dot hybridization,
$\Gamma=\Gamma_L+\Gamma_R$,  $\mu_L=-\mu_R=0.2$ as a symmetric bias.
The reservoirs are modeled by flat bands with a sharp cutoff at
$D=\pm 1$, including $N=501$ electronic states for each reservoir.}
\label{FigNI}
\end{figure}

\begin{figure}[htbp]
\vspace{0mm} \hspace{0mm} {\hbox{\epsfxsize=85mm \epsffile{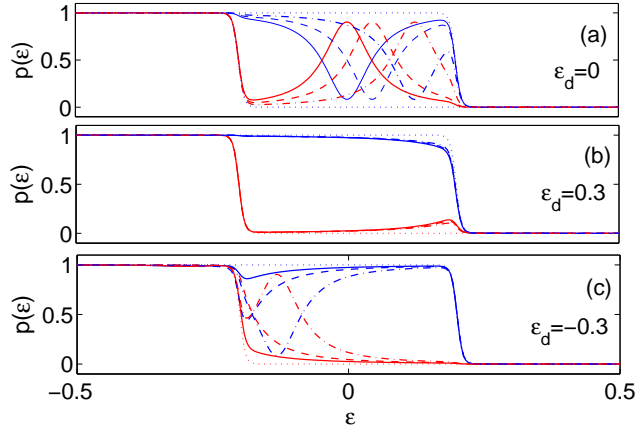}}}
\caption{Population of reservoirs' states from a mean-field QLE
treatment. Plotted are the $L$ (three top lines) and $R$ (three
bottom lines) occupations as a function of electron energy at
$\Gamma t=15$ using the same set of parameters as in Fig.
\ref{FigNI}. (a) $\epsilon_d=0$, (b) $\epsilon_d=0.3$, and (c)
$\epsilon_d=-0.3$. Full, dashed, and dashed-dotted lines correspond
to $U=0, 0.1, 0.3$, respectively. The initial distributions for both
the $L$ and $R$ reservoirs are presented by dotted lines. }
\label{FigMF}
\end{figure}


\section{Applications}
\label{cd}

We now turn our attention to applications of the preceding methods.
We first study the effects of Coulombic interactions on the
reservoirs' DOF evolution. We later investigate the equilibration
process in the system, through the implementation of probes.

\subsection{Anderson model with electron-electron interactions}
\label{ee}

In this section we study the evolution of the finite-size Anderson
model with or without interactions, based on the three methods
described earlier in Sec. \ref{methods}. As mentioned above, these
techniques provide the dynamics of the total DM. While the
fingerprints of many-body effects are disguised in the time
evolution of conventional quantities, e.g., in the dot occupation
and the charge current, they are well manifested in the reservoirs'
population dynamics, allowing us to discern microscopic many-body
scattering processes from single-particle events.


The population $p(\epsilon_k)=\langle c_k^{\dagger}c_k\rangle$ of
both reservoirs, in the {\it noninteracting} case, is displayed in
Fig. \ref{FigNI} at different times. We note that results obtained
using the QLE framework of Sec. \ref{sec-qle} perfectly agree with
numerically-exact fermionic trace formula simulations. The three
panels present results using different values for the dot energy.
(a) When the dot energy is placed within the bias-window
($\epsilon_d=0$) a resonance feature develops around the position of
the dot level, with a dip (peak) showing in the $L$ ($R$) bath. In
contrast, if the dot energy is positioned either above the bias
window (b) or below it (c), a dot-assisted tunneling feature
develops, with population transfer taking place around available
states that are the nearest in energy to $\epsilon_d$. The dynamics
shown in Fig. \ref{FigNI} is reversible, with a characteristic time
$\tau_{d}\sim 2\pi/\Delta E$, $\Delta E=2D/N$ is the mean spacing
between energy levels and $N=N_{L,R}$ is the number of states in the
$L,R$ baths  \cite{yuka}. At this characteristic time the dot
population begins to vary from its QSS value due to finite size
effects. This behavior can be captured with trace formula
simulations, but not within the QLE
approach. 

In Fig. \ref{FigMF} we display the dynamics under a mean-field QLE
treatment with parameters corresponding to Fig. \ref{FigNI}. While
we are mindful of the technique's known pathologies \cite{BerMF}, we
stress that this calculation provides an intuitive understanding of
the role of interactions: Within MF, the effect of finite $U$ is to
shift features in concert with the renormalized dot energy,
$\tilde{\epsilon}_{d}(t)=\epsilon_{d}+U\left\langle
n_{d}(t)\right\rangle$ [Eq. (\ref{eq:edtilde})]. Interestingly,
panel (c) demonstrates a change in transport mechanism, from a
dot-assisted tunneling at small $U$, to resonance transmission at
large $U$, since the renormalized dot energy enters the bias window
at a large enough interaction strength. Therefore, e-e
interactions can enhance or suppress electronic transport, depending
on the dot bare energy position.

\begin{figure}[htbp]
\vspace{0mm} \hspace{0mm} {\hbox{\epsfxsize=80mm \epsffile{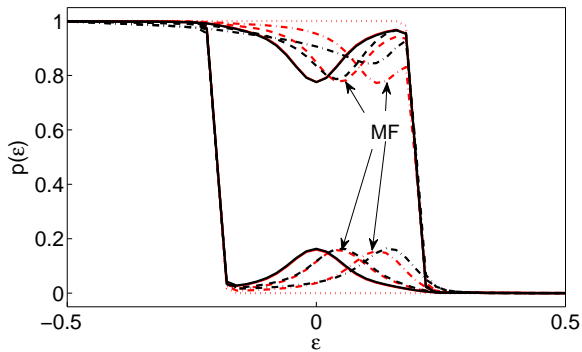}}}
\caption{Simulations of reservoirs' levels population in the
resonance case, $\epsilon_d=0$, at $\Gamma t=4$, using INFPI (dark
curves) and MF QLE (light curves). Full, dashed and dashed-dotted
lines were obtained using $U=0, 0.1, 0.3$, respectively. Top lines
correspond to the $L$ bath distribution, the bottom lines to the $R$
bath. The initial distributions of both $L$ and $R$ reservoirs are
presented by dotted lines. Parameters are the same as in Fig.
\ref{FigNI}, with $N_{L,R}=101$ bath states. INFPI numerical
parameters are $\delta t=1$ and $N_s=7$. } \label{FigINFPI1}
\end{figure}

Mean-field results are compared to numerically-exact INFPI
simulations in Fig. \ref{FigINFPI1} for $U=0.1$ and $U=0.3$, with
the bare dot energy centered within the bias window. Data was
produced by time evolution of all $\langle c_k^{\dagger}c_k\rangle$,
$k=l,r$, expectation values up to $\Gamma t\sim4$. In agreement with
MF QLE results, the basic effect of e-e interactions observed here
is a shift in the resonance position. Overall, we conclude that MF
simulations can reproduce the dynamics for this set of parameters,
up to $U/\Gamma\lesssim2$. Qualitative features are correct through
$U/\Gamma\sim6$.

\begin{figure}[htbp]
\vspace{0mm} \hspace{0mm} {\hbox{\epsfxsize=80mm \epsffile{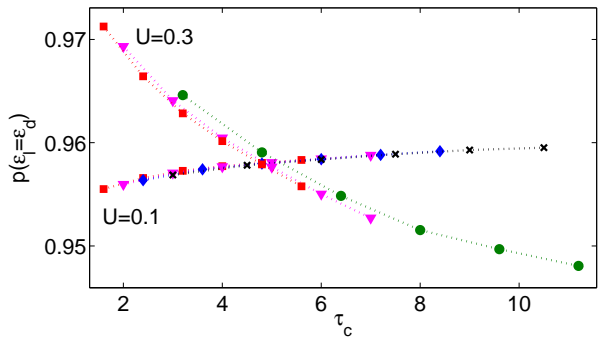}}}
\caption{Convergence trend for the INFPI data of Fig.
\ref{FigINFPI1}. We plot $p(\epsilon_l=\epsilon_d)$ as a function of
$\tau_c$ at $\Gamma t=1.2$ for $U=0.1$ and $U=0.3$. Results are
shown using different time steps, $\delta t=0.8$ ($\square$), $\delta
t=1.0$ ($\triangledown$),  $\delta t=1.2$ ($\lozenge$), $\delta t=1.5$
($\times$) and $\delta t=1.6$ ($\circ$). Other parameters are the
same as in Fig. \ref{FigINFPI1}.} \label{FigINFPI2}
\end{figure}

Convergence of INFPI is verified with respect to the time step
adopted, $\delta t$, and the memory time accounted for, $\tau_c$.
Representative convergence curves for $p(\epsilon_l=\epsilon_d)$ are
depicted in Fig. \ref{FigINFPI2}. While the time-step used does not
affect our results, we note that the data is not yet fully converged
with respect to $\tau_c$. This slow convergence could be attributed
to the long decorrelation time experienced by an electron residing
on any particular bath level, since its decorrelation process should
take place by following a two-step procedure: the electron should
first leave the particular bath state and populate the quantum dot.
From the dot, it may subsequently transfer to any other bath state. One should
also note that we display here a convergence curve for a particular
level. It is more accurate, but computationally demanding, to look
at the overall evolution of $p(\epsilon)$ (for all $\epsilon$) with
$\tau_c$. This is because the resonant feature may change its
magnitude with $\tau_c$, as we see here, as well as its absolute
position. Fig. \ref{FigINFPI2} only analyzes the effect of $\tau_c$
on the peak magnitude. 


\begin{figure}[htbp]
\vspace{0mm} \hspace{-5mm} \rotatebox{-90}{  \hbox{\epsfxsize=80mm
\epsffile{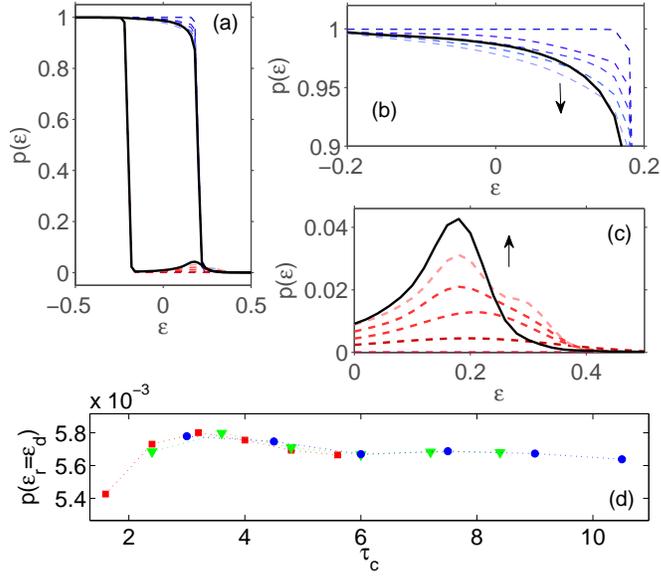}}} \caption{(a) INFPI data for the reservoirs
population at different times for $\epsilon_d=0.25$ and $U=0.1$. The
five top (bottom) lines depict the population of the $L$ ($R$) baths
at times $\Gamma t=$ 0.05, 1, 2, 3, and 4.
(b) Zooming over the $L$ bath population.
(c) Zooming over the $R$ bath population.
The arrows indicate on the direction of time evolution.
In  (a)-(c) the full lines represent MF QLE data
at the latest time, $\Gamma t=4$.
(d) Convergence behavior of $p(\epsilon_r=\epsilon_d)$
at $\Gamma t=1.2$ using different time
steps, $\delta t=0.8$ ($\square$), $\delta t=1$ ($\triangledown$) and $\delta
t=1.5$ ($\circ$). Other parameters are the same as in Fig.
\ref{FigNI}, with $N_{L,R}=101$ bath states.
 } \label{FigINFPI3}
\end{figure}

In Fig. \ref{FigINFPI3} we study the dynamics of the reservoirs'
levels population when the renormalized dot energy sits above the
bias window, $\tilde \epsilon_d > \mu_L$. The bare energy is taken
at $\epsilon_d=0.25$ and the interaction strength is $U=0.1$. By
separately calculating the time evolution of the dot occupation, to
produce $\langle n_{d,\sigma}\rangle \sim$ 0.12 in the QSS limit
(valid for $\Gamma t\gtrsim2$), we estimate the renormalized dot
energy to be about $\tilde \epsilon_d\sim 0.26$. Overall,
occupations change very slightly in time, since the dot is
off-resonant thus the transport follows a dot-assisted tunneling
mechanism. The reservoirs' dynamics still clearly manifests
many-body effects that are not included in a MF (effective
single-body) description. Essentially, we find that electrons
populate high energy levels in the $L$ and $R$ baths up to
$\epsilon_k\sim 0.45$. This high-energy population cannot be
explained by the dot-level shift or the dot finite broadening
$\Gamma$, as this could account for population up to $\epsilon_k\sim
0.35$ only, see  MF data (full line) in panel (c). It should be
noted that the population of levels that are initially empty,
$p(\epsilon>\mu_L)$, develops identically at the $L$ and $R$
reservoirs. In other words, the high energy tails in panel (c)
represent the occupation of states both in the $L$ and $R$ baths.
Supporting convergence behavior is included in panel (d). We have
also performed simulations when taking a stronger interaction,
$U=0.3$ and $\epsilon_d=0.15$, to yield $\tilde \epsilon\sim 0.21$.
In this case the population shows a high-energy tail that is further
enhanced with respect to MF data, representing significant
deviations from a single particle description. However, as we did
not manage to fully converge these results, they are not included
here.

\begin{figure}[htbp]
\vspace{0mm} \hspace{0mm} {\hbox{\epsfxsize=85mm
\epsffile{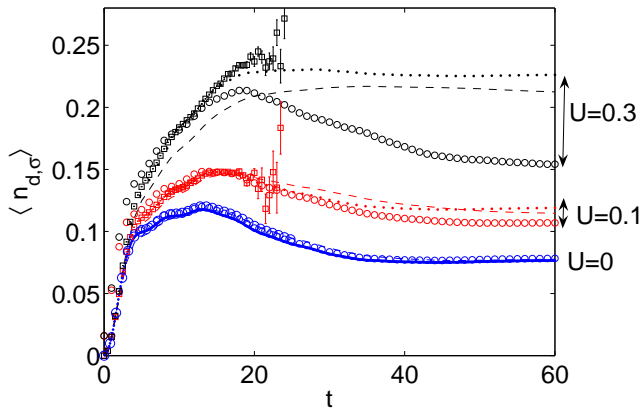}}}\caption{Dot occupation as a function of time,
generated from four different methods: INFPI (dotted), Monte-Carlo
simulations ($\square$) \cite{MC1,MC2,MC3,MC4},
MF QLE equations (dashed) and perturbation
theory treatment ($\circ$) \cite{Komnik}. Results are shown for three setups,
bottom to top: $\epsilon_d=0.3$ with $U=0$, $\epsilon_d=0.25$ with
$U=0.1$, and $\epsilon_d=0.15$ with $U=0.3$. Other parameters are
the same as in Fig. \ref{FigNI}, with $\delta t=1$, $N_s=7$ and $N_{L,R}=101$ bath states.}
\label{FigINFPI5}
\end{figure}

The breakdown of the single-particle picture is difficult to discern
in cumulative quantities such as the charge current and the dot
occupation, the latter is presented in Fig. \ref{FigINFPI5}, where
we follow the time evolution of this quantity using four different
techniques: MF QLE equations, first-order perturbation theory method
\cite{Komnik}, Monte-Carlo simulations \cite{MC1,MC2,MC3,MC4} and
INFPI \cite{dvira_prb}. The comparison shows a very good agreement
between the latter two exact methods up to $\Gamma t\sim 1$
\cite{dvira_prb}. MF QLE theory and first-order perturbation
expansion both predict the correct behavior at $U/\Gamma\lesssim 2$.
At strong interactions ($U/\Gamma\sim 6$), perturbation theory
fails, while MF QLE equations still provides correct qualitative
behavior for $\langle n_{d,\sigma}\rangle$. Note that we implement a
sharp cutoff at $D=\pm 1$ in all methods. Since we do not account
for the (small) energy shift in Eq. (\ref{eq:cdQLE}), MF QLE data
suffer from a small shift in values, see also Fig. \ref{Figtau}.

Naively considering the dot occupation only, one may conclude that at
$U/\Gamma\lesssim 6$ many-body effects are contained in a
mean-field, effective single-body, description. However, traces of
energy resolved reservoirs' dynamics as we show in Fig.
\ref{FigINFPI3} expose the existence of interaction effects beyond
mean-field, resulting in levels population beyond the resonance
width. The dot occupation thus {\it withholds} mechanisms involved
in the transport process, while detailed reservoirs' level
population can illuminate extant many-body effects and their
energy resolution.


We now comment on the simulation time of a convergence analysis as
presented in Fig. \ref{FigINFPI2}, covering three different time
steps and $N_s=2,...,7$. Convergence should in principle be verified
for all bath states ($N_L+N_R=101\times 2$) and at all times. In
practice, we have tested it for representative states only. The
MATLAB implementation of the computational algorithm took advantage
of the MATLAB built-in multi-threaded parallel features and utilized
100$\%$ of all available CPU cores on a node. When executed on one
cluster node with two quad-core 2.2GHz AMD Opteron cpus and 16GB
memory, convergence analysis for each expectation value took about
7x24 hours and 250MB of memory. Computations performed on the GPC
supercomputer at the SciNet HPC Consortium \cite{SciNet} were three
times faster. Computational time scales linearly with the simulated
time $t$. For a fixed $N_s$ value, the computational effort does not
depend on the system temperature and the value of $U$ employed.


\subsection{Quantum equilibration and thermalization}
\label{dcd}

The techniques developed in Sec. \ref{methods} provide the time
evolution of the total DM of a large system, allowing us to address
next the problem of equilibration and thermalization in quantum
mechanics. The basic question of interest here is how do quantum
systems equilibrate from a nonequilibrium initial preparation, if at
all. Furthermore, it is of importance to understand under what
conditions a system may approach one of the canonical ensembles of
statistical mechanics.

\begin{figure}[htbp]
\vspace{0mm} \hspace{0mm} {\hbox{\epsfxsize=75mm
\epsffile{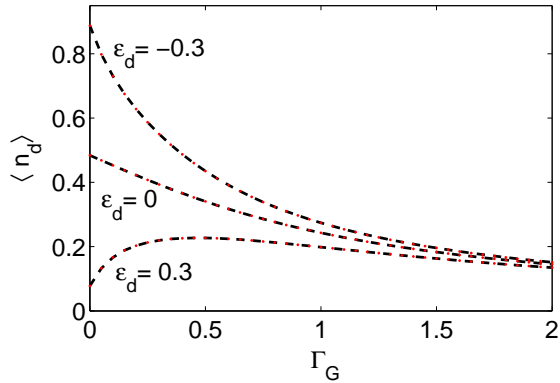}}} \caption{Steady-state dot occupation under a
voltage probe condition (dashed) and a dephasing probe (dot) for
different values of the dot energy, $\epsilon_d=-0.3$, 0 and $0.3$,
top to bottom. Data was obtained by simulating Eq.
(\ref{eq:dot-occ}) in the long time limit using $\mu_L=-\mu_R=0.2$,
$\beta=200$ and $\Gamma_{L,R}=0.025$.} \label{FigD}
\end{figure}

Before addressing the equilibration problem in detail we present in
Fig. \ref{FigD} a more standard quantity, the steady-state dot
occupation. This will serve us for motivating the study of the total
DM for resolving transport mechanisms. The dot occupation is
displayed here as a function of $\Gamma_G$, the dot-probe
hybridization strength, and we show results using either a voltage
probe or a dephasing probe, for different values of the dot energy
position. We find that the dot occupation is insensitive to the
probe condition, an observation that can be proved analytically by
studying the long time behavior of Eq. (\ref{eq:dot-occ}) under
either probes, Eqs. (\ref{eq:voltage}) or (\ref{eq:dephasing}). It
is interesting to note the crossover behavior of the dot occupation
when its energy is placed above the bias window: When
$\Gamma_G<\Gamma_{L,R}$ occupation grows linearly with $\Gamma_G$.
However, it decays as $\Gamma_G^{-1}$ at large values, when
effective dephasing and inelastic effects are strong. This behavior
is similar to the thermally assisted tunneling behavior observed
when using more detailed modeling \cite{DRed}. It can similarly be
shown that the steady-state current in the system is identical
irrespective of the probe condition. Note that we restrict ourselves
to the quasi steady-state region since the $G$ bath does not serve
as a proper probe before quasi steady-state sets in.

The underlying transport mechanisms are therefore {\it obscured} in
cumulative quantities (current, occupation) in this probe model, as
well as in more explicit electron-phonon modeling \cite{Nitzanrev}.
Details about the involved mechanisms can be resolved by studying,
e.g. the current noise, inelastic electron tunneling spectra
\cite{Nitzanrev}, and the evolution of the reservoirs population
\cite{DviraP}, as we show below in the context of quantum
equilibration and thermalization.

The equilibration problem in quantum mechanics could be considered
within different setups: a closed system \cite{Reimann1,
Reimann,ShortNJP}, a system-bath scenario \cite{short,h_pra}, or taking
peer quantum systems \cite{h2}.
Here, we consider the Anderson model with a probe, excluding e-e
interactions, and simulate the following setup: At time $t=t_0$ we
put into contact through a quantum dot two reservoirs each
separately prepared in grand canonical states at chemical potentials
$\mu_{\nu}$ and temperature $\beta^{-1}$. Electrons on the dot are
susceptible to either elastic-decoherring processes or inelastic
effects, mimicked by coupling to the relevant probe. For a schematic
representation, see Fig. \ref{FigS}(b). Given this scenario, we
investigate whether the two reservoirs can equilibrate or even
thermalize in time, and furthermore, the nature of the equilibrium
state. As we show below, when only elastic dephasing effects are
mimicked, the system approaches a non-canonical equilibrium state.
When inelastic processes are emulated, the two reservoirs relax
towards a common canonical state. It should be noted that these
results can be obtained for a {\it finite and closed} system, under
a unitary evolution \cite{qle}.


\begin{figure}[htbp]
\hspace{2mm} {\hbox{\epsfxsize=80mm \epsffile{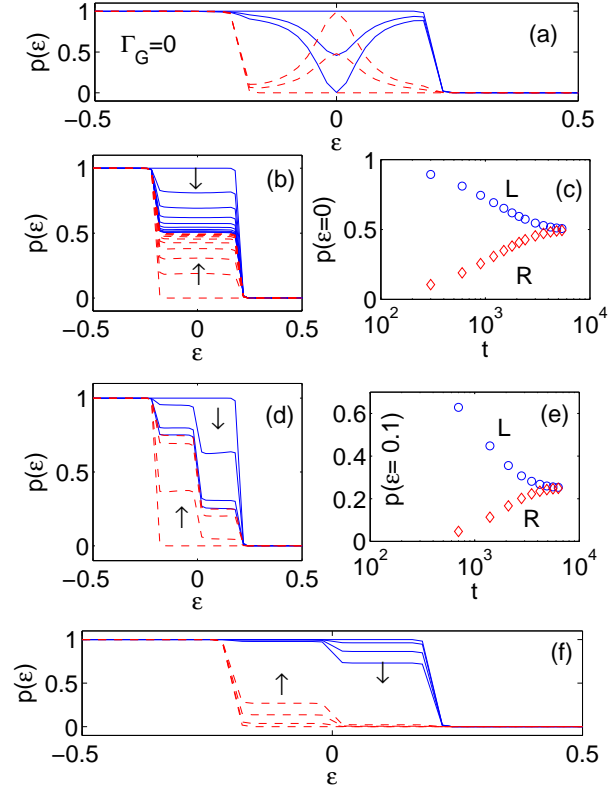}}} \caption{
Dynamics of the reservoirs population in the Anderson probe model.
(a) Population of the $L$ (full) and $R$ (dashed) reservoirs,
$\Gamma_G=0$.  We show results at $t=0$, $180$, $340$ as the
resonance pattern develops in time.
(b)-(c) Equilibration with a dephasing probe, $\Gamma_G=0.4$,
approaching a non-canonical equilibrium state. The $L$ (full) and
$R$ (dashed) population are shown at $t=0,\delta t,...,9\delta t$,
$\delta t=600$. Panel (c) demonstrates a slow-down in dynamics in
approaching the equilibrium state.
(d)-(f) Approaching thermal equilibrium with a voltage probe. In
panel (d) $\Gamma_G=0.4$ with the $L$ (full) and $R$ dashed line
population shown at $t=0, 700, 2800, 6300$. Panel (e) displays the
population as a function of time at a certain energy,
$\epsilon=0.1$. Panel (f) presents information as in (d), with
$\Gamma_G=5$. In all panels $\beta=200$, $\Gamma_L=\Gamma_R=0.025$,
$\epsilon_d=0$, $\mu_L=-\mu_R=0.2$, $D=1$, $N_{L,R}=101$ and
$N_G=2001$. The arrows mark the direction of propagation in time.}
\label{FigG1}
\end{figure}


We identify thermal equilibration in our peer quantum system setup,
by adjusting the conditions of Refs. \cite{short,gogolin}, demanding
that: (i) The system should equilibrate, i.e., evolve towards some
particular state, and stay close to it for almost all time.
Furthermore, the equilibrium state should be (ii) independent of the
dot properties-energetics and initial state, (iii) insensitive to
the precise initial state of each reservoir, (iv) close to diagonal
in the energy basis of its eigen-Hamiltonian, and (v) a canonical state.

We use the trace formula, an exact unitary method, and follow the
reservoirs' mutual equilibration process. We evolve the system 
using either a dephasing probe or a voltage probe, see Fig. \ref{FigG1},
up to the time where recurrence features start to manifest, found here to
scale as $\tau_{rec}\propto \sum_{i=L,R,G}N_i$.
As a reference, panel (a) displays results for the model without a probe,
showing the development of a resonance feature around the dot energy
position at $\epsilon_d=0$. A clear evolution towards an equilibrium
state is demonstrated when a probe is presented. With a dephasing
probe, (b)-(c), the population of the two reservoirs relax to a
two-step function with $p(\mu_R<\epsilon<\mu_L)\sim0.5$. Because
electrons from the $L$ grain loose their phase memory on the dot,
half populate the $R$ side, on average, in the long time limit. This
equilibrium state is sensitive to the precise details of the initial
electron distribution, as energy redistribution is not allowed. We
build a large $G$ to delay recurrence behavior, but note that
results at earlier times do not depend on the size of $G$,
reinforcing the observation that $G$ acts as an agent in driving the
$L$-$R$ mutual equilibration. When inelastic effects are mimicked
with a voltage probe, and $\Gamma_G$ is large enough, panel (f), the
system approaches a Gibbs-like thermal state--- a step function at
zero temperature. Results are shown up to the time $\tau_{rec}$ at
which recurrence features develop, which emerges here {\it before}
full thermalization takes place. In order to achieve full
thermalization one should further increase the size of the $G$ bath,
so as to delay recurrences. Alternatively, a dissipative mechanism
could be introduced into $G$, e.g. by building a hierarchy of its
interactions with the $L$-$R$ system. Using a smaller value for
$\Gamma_G$, a non-canonical equilibrium distribution develops
(d)-(e), reflecting the contribution of coherent and (effectively)
incoherent electrons in the dynamics. It is also interesting to
compare panels (c) and (e), displaying the equilibration progress
for a dephasing probe and a voltage probe, respectively, while
maintaining the value of $\Gamma_G$. We find that the characteristic
timescale to reach equilibrium is very similar in both cases. Thus,
while the probe type dictates the structure of the equilibrium
state, it does not affect the equilibration timescale.

\begin{figure}[htbp]
\vspace{1mm} \hspace{0mm} {\hbox{\epsfxsize=80mm \epsffile{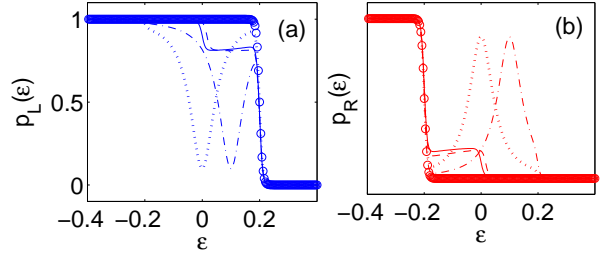}}}
\caption{(a) Occupation of $L$ bath at time $t=0$ ($\circ$) and
$t=1500$ for $\Gamma_G=0$ and $\epsilon_d=0$ (dotted), $\Gamma_G=0$
and $\epsilon_d=0.1$ (dashed-dotted), $\Gamma_G=0.4$ and
$\epsilon_d=0$ (full), $\Gamma_G=0.4$ and $\epsilon_d=0.1$ (dashed).
The latter two lines assume the voltage probe condition.
(b) Same for the $R$ side occupations. Other parameters are the same
as in Fig. \ref{FigG1}. } \label{FigG2}
\end{figure}

Fig. \ref{FigG2} shows that while under coherent evolution the
resonance peak emerges around the energy $\epsilon_d$, in the
presence of a voltage probe with (large enough) $\Gamma_G$, the
buildup of the equilibrium state systematically occurs around the
equilibrium Fermi energy. This holds even when the dot is placed
{\it outside} the bias window (not shown). Analogous trends take
place when allowing for dephasing only.

\begin{figure}[htbp]
\vspace{0mm}  {\hspace{-8mm} \hbox{\epsfxsize=90mm \epsffile{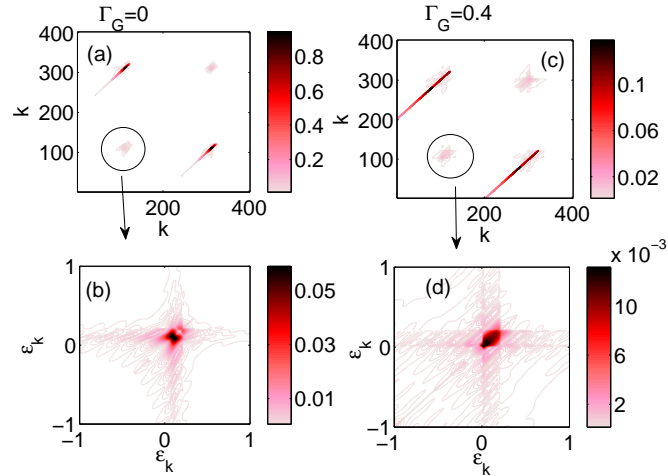}}}
\caption{Absolute values of the density matrix elements
$\rho_{k,k'}$, $k,k'=l,r$, at $t=600$, excluding diagonal elements. 
In panel (a) and (b) $\Gamma_G=0$. In panels (c) and (d) we use a voltage
probe with $\Gamma_G=0.4$.
Panels (a) and (c) display the total density matrix, and the axes are the energy indices
1,...,402. The $L$ reservoir includes the first 201 states. The rest are $R$ bath states.
Panels (b) and (d) zoom on the $\rho_{l,l'}$ density matrix, the bottom-leftmost part of the
total DM.
Other parameters are
$\epsilon_d=0.1$, $\Gamma_{L,R}=0.025$, $\beta=200$, $\mu_L=-\mu_R=0.2$.
} \label{FigDM}
\end{figure}

A thermal equilibrium state should be diagonal in the energy
eigenbasis of its Hamiltonian \cite{short}. In Fig. \ref{FigDM} we
display the density matrix $\rho_{k,k'}=\langle c_k^{\dagger}c_{k'}
\rangle$, $k,k'=l,r$, excluding diagonal elements $\rho_{k,k}$,
without a voltage probe (a)-(b), and with a one (c)-(d), using the
QLE technique. This quantity is expected to oscillate in the long
time limit since the Hamiltonian is not diagonal in the (local) $l$
and $r$ bases. We still show the results in the local reservoirs'
basis, so as to manifest local $\nu$-bath properties. There are
three significant differences in the behavior of off-diagonal
elements, with and without the probe: (i) The absolute value of the
coherences, at a given time, is smaller when $\Gamma_G\neq 0$. (ii)
The DM approaches a diagonal form (strict diagonal values are not
shown). (iii) When $\Gamma_G=0$, oscillations occur around the dot
energy position $\epsilon_d$. With the probe, contributions are
scattered, yet they appear more prominently around the equilibrium Fermi
energy, $\epsilon=0$. These three features should become
more pronounced at longer times, which can be simulated using the trace formula
approach.


\section{Conclusion}
\label{conclusions}

We have extended analytical and numerical methods, developed for
simulating the dynamics of impurities, i.e., subsystems attached to
large reservoirs, to reveal the dynamics of the total system. As an
example, we have focused on the Anderson model, a quantum dot
coupled to two metal grains, and obtained the evolution of the total
density matrix, focusing on the reservoirs' evolution from an
initial nonequilibrium state. We have studied the noninteracting
model, as well as a model with interactions and a probe model,
emulating elastic dephasing and dissipation effects. The three
methods presented are the analytic quantum Langevin equation
approach, a simulation based on a trace formula, and an exact
numerical path integral scheme that can accommodate e-e repulsion
effects. Notably, the extension of the QLE treatment to provide the
total DM is of general importance as it can be used in multitude of
other systems, as long as one can identify a ``subsystem" within the
total system.

Making use of the methods developed, we have investigated the total
system dynamics in the presence of distinct effects: (i) e-e
interactions on the impurity, and (ii) dephasing and inelastic
scattering effects. Addressing the prior, our calculations allow us
to {\it energy resolve} the effect of e-e interactions on electron
transfer in the Anderson dot model. In the resonant regime we found
that the dynamics observed for noninteracting electrons is largely
preserved up to $U/\Gamma\lesssim 2$, and the main effect of
interactions on the reservoirs' occupation is apparently a simple
shift in the position of features affected by the renormalization of
the dot energy. Away from resonance, in the tunneling domain, the
presence of weak interactions already manifested itself in
scattering electrons to high energy levels, an effect that is not
captured within a mean-field treatment. In the case of the later
effect, we found numerically that the presence of dephasing and
inelastic effects on a weak link only can lead to {\it global}
system equilibration and even thermalization. It is important to
note that no restrictions were enforced on the metals' band
structure and the dot energy. This is significant in light of many
other studies in which equilibration requires the ``nondegenerate
energy gap" condition to be satisfied \cite{Reimann1,short,h2}.

Future directions include the study of finite temperature and
electron-electron interaction effects in the equilibration process
\cite{dvira_prb}, and the behavior given a quantum dot chain between
the two metal grains. In a linear chain of impurities we expect that
the coherent-diffusive crossover in the charge current behavior
\cite{Lebowitz} would similarly  manifest itself in the energy
reorganization process of the reservoirs. The methods developed here
could also be adopted for the study of bosonic systems, e.g., to
describe the dynamics of bosonic degrees of freedom interacting with
harmonic baths.


\begin{acknowledgments}
This work has been supported by an NSERC discovery grant. M.K.
thanks Diptiman Sen for useful discussions.
The authors acknowledge
T. L. Schmidt for providing the time-dependent perturbation theory
code.
The work of K.L.T. has
been supported by an Early Research Award of D.S.
Computations were performed on the GPC supercomputer
at the SciNet HPC Consortium \cite{SciNet}. SciNet is funded by:
the Canada Foundation for Innovation under the auspices of Compute Canada;
the Government of Ontario; Ontario Research Fund - Research Excellence;
and the University of Toronto.
\end{acknowledgments}


\renewcommand{\theequation}{A\arabic{equation}}
\setcounter{equation}{0}  
\section*{Appendix A: Density matrix elements in the quantum Langevin approach}

We provide here the explicit expressions for the density matrix
elements $\rho_{l,k}$, $k\in L,R$. The population is calculated by
evaluating the correlation functions in Eq. (\ref{eq:pl}), using
Eqs. (\ref{eq:clm}) and (\ref{eq:cdi}), to yield
\bea &&p(\epsilon_l)= \langle c_l^{\dagger}(t_0)c_l(t_0)\rangle
+F_2+F_3  \eea
The first term accommodates the initial condition. The second
($F_2$)  and the third ($F_3$) terms are given by
\bea F_{2}&=&-v_{l}^{2}f_{l} \frac{2\Gamma }{\Gamma^{2}+
\epsilon_{dl}^{2}} t -2v_{l}^{2}f_{l}\frac{\epsilon_{dl}^{2}
-\Gamma^{2}}{\left(\epsilon_{dl}^{2}+ \Gamma^{2}\right)^{2}}
+\frac{v_{l}^{2}f_{l}e^{-\Gamma
t}}{\left(\epsilon_{dl}^{2}+\Gamma^{2}\right)^{2}}\Biggl
\{2\left(\epsilon_{dl}^{2}-\Gamma^{2}\right)\cos\left(\epsilon_{dl}t\right)
+ 4\epsilon_{dl}\Gamma\sin\left(\epsilon_{dl}t\right)\Biggr\}
\nonumber\\
F_{3}&=&v_{l}{}^{2}\sum_{k=l,r} \frac{v_{k}{}^{2}f_{k}}
{\Gamma^{2}+\epsilon_{dk}^{2}}\Biggl\{\frac{4\sin^{2}\left
(\frac{\epsilon_{lk}}{2}t\right)}{\epsilon_{lk}^{2}}
+\frac{1}{\Gamma^{2}+\epsilon_{dl}^{2}}\left[e^{-2\Gamma t}+1-e^{t
(i\epsilon_{dl}-\Gamma)}-e^{-t(i\epsilon_{dl}+\Gamma)}\right]
\nonumber\\
&+&\left[\frac{1-e^{-t(\Gamma+i\epsilon_{dl})} +
 e^{-t(\Gamma+i\epsilon_{dk})}
-e^{-it\epsilon_{lk}}} {\left(\epsilon_{dl}
-i\Gamma\right)\epsilon_{lk}} +c.c.\right]\Biggr\}.
\label{eq:F2F3} \eea
%
Inter and intra-reservoir coherences, e.g. $\rho_{l,k}(t)$, $k\in L,
R$ can be similarly calculated. Here, one should distinguish between
the cases  $\epsilon_l=\epsilon_k$ and $\epsilon_{l} \neq
\epsilon_{k}$. In the latter case  we find that
\bea \rho_{l,k}(t)\equiv \left\langle
c_{l}^{\dagger}(t)c_{k}(t)\right\rangle =A_{1}+A_{2}+A_{3}
\label{eq:non-dia} \eea
where \bea
A_{1}&=&-\frac{v_{k}v_{l}f_{l}}{\Gamma+i\epsilon_{dl}}\left[\frac{e^{-t(\Gamma+i\epsilon_{dl})}-
e^{i\epsilon_{lk}t}}{\Gamma+i\epsilon_{dk}}+\frac{i}{\epsilon_{lk}}
(1-e^{i\epsilon_{lk}t})\right]
\nonumber\\
A_{2}&=&-\frac{v_{l}v_{k}f_{k}}{\Gamma-i\epsilon_{dk}}\left
[\frac{e^{-t(\Gamma-i\epsilon_{dk})}-e^{i\epsilon_{lk}t}}{\Gamma-i\epsilon_{dl}}+
\frac{i}{\epsilon_{lk}}(1-e^{i\epsilon_{lk}t})\right] \nonumber
\eea and
%
\bea A_{3}&=&
 v_{l}v_{k}\sum_{k^{\prime}\in L,R}\frac{v_{k^{\prime}}^{2}f_{k^{\prime}}}
{\Gamma^{2}+\epsilon_{dk^{\prime}}^{2}}
\nonumber\\
&\times& \Biggl\{\frac{1}{\epsilon_{k^{\prime}l}
\epsilon_{k^{\prime}k}} +\frac{e^{it\epsilon_{lk^{\prime}}}}
{\epsilon_{lk^{\prime}}\epsilon_{k^{\prime}k}}
-\frac{e^{it\epsilon_{lk}}}{\epsilon_{lk^{\prime}}
\epsilon_{k^{\prime}k}} -\frac{e^{it\epsilon_{k^{\prime}k}}}
{\epsilon_{k^{\prime}l}\epsilon_{k^{\prime}k}} +
\frac{e^{it\epsilon_{k^{\prime}k}}+e^{-t(\Gamma+i\epsilon_{dl})}
-e^{it\epsilon_{lk}}-e^{-t(\Gamma+i\epsilon_{dk^{\prime}})}}
{\epsilon_{lk^{\prime}}(i\Gamma+\epsilon_{k d})}
\nonumber\\
&+&\frac{e^{it\epsilon_{lk^{\prime}}} +
e^{-t(\Gamma-i\epsilon_{dk})} -e^{it\epsilon_{lk}}-
e^{-t(\Gamma-i\epsilon_{dk^{\prime}})}}{\epsilon_{k^{\prime}k}
(i\Gamma+\epsilon_{dl})} +\frac{e^{-2\Gamma t}+e^{it\epsilon_{lk}}
-e^{-t(\Gamma+i\epsilon_{dl})}-e^{-t(\Gamma-i\epsilon_{dk})}}{(i\Gamma+\epsilon_{dl})
(-i\Gamma+\epsilon_{dk})}\Biggr\}.
\nonumber \eea
%
In the resonant limit, $\epsilon_{l} = \epsilon_{k}$ and $k\notin
L$, a simpler result is obtained,
\bea \left\langle c_{l}^{\dagger}(t)c_{k}(t)\right\rangle
=A_{1}^{r}+A_{2}^{r}+A_{3}^{r} \label{eq:diag-like} \eea
with \bea A_{1}^{r}&=&-\frac{v_{k}v_{l}f_{l}}{\Gamma+i\epsilon_{dl}}
\left[t-\frac{1-e^{-t\left(\Gamma+i\epsilon_{dl}
\right)}}{\Gamma+i\epsilon_{dl}}\right]
\nonumber\\
A_{2}^{r}&=&-\frac{v_{l}v_{k}f_{k}}{\Gamma-i\epsilon_{dl}}
\left[t-\frac{1-e^{-t\left(\Gamma-i\epsilon_{dl}\right)}}{\Gamma-i\epsilon_{dl}}\right]
\nonumber\\
A_{3}^{r}&=&\frac{v_{k}}{v_{l}}F_{3}.
\nonumber
\eea
%
\bibliographystyle{iopart-num.bst}

\bibliography{Impurity_References}

\begin{thebibliography}{10}
\bibitem{comm}
Eq. (\ref{eq:igA}) can be justified by repeating the steps leading to Eq.
  (\ref{eq:im_der}), and taking the long-time limit
\bibitem{foot_dense}
For dense reservoirs, one could replace the summation by an integral over
  frequency. However, since we are simulating the dynamics using finite number
  of bath states, we prefer to show our results in its discrete form
\end{thebibliography}

\end{document}